\documentclass[aps,twocolumn,superscriptaddress,nofootinbib]{revtex4}

\usepackage{amsmath,amssymb,graphicx}
\usepackage{color}
\usepackage[normalem]{ulem}

\graphicspath{{fig-ps/}}

\definecolor{yblue}{rgb}{0.06, 0.3, 0.57}

\begin{document}

\title{Vortex precession dynamics in general radially symmetric potential traps
  in two-dimensional atomic Bose-Einstein condensates}

\author{P. G. Kevrekidis}
\affiliation{Department of Mathematics and Statistics, University of Massachusetts,
Amherst, Massachusetts 01003-4515 USA}

\author{Wenlong Wang}
\email{wenlongcmp@gmail.com}
\affiliation{Department of Physics and Astronomy, Texas A$\&$M University,
College Station, Texas 77843-4242, USA}

\author{R. Carretero-Gonz{\'a}lez}
\affiliation{Nonlinear Dynamical Systems
Group,\footnote{\texttt{URL}: http://nlds.sdsu.edu}
Computational Sciences Research Center, and
Department of Mathematics and Statistics,
San Diego State University, San Diego, California 92182-7720, USA}

\author{D. J. Frantzeskakis}
\affiliation{Department of Physics, National and Kapodistrian University of Athens,
Panepistimiopolis, Zografos, 15784 Athens, Greece}

\author{Shuangquan Xie}
\affiliation{Department of Mathematics and Statistics, Dalhousie University, Halifax, Canada}

\begin{abstract}
We consider the motion of individual two-dimensional vortices in 
general radially symmetric potentials in Bose-Einstein condensates. 
We find that although in the special
case of the parabolic trap there is a logarithmic correction
in the dependence of the precession frequency $\omega$
on the chemical potential $\mu$,
this is no longer true for a general potential $V(r) \propto r^p$.
Our calculations suggest that for $p>2$, the 
precession frequency scales with $\mu$ as  
$\omega \sim \mu^{-2/p}$. This theoretical prediction is corroborated
by numerical computations, both at the level of spectral (Bogolyubov-de
Gennes) stability analysis by identifying the relevant
precession mode dependence on $\mu$, but also through direct
numerical computations of the vortex evolution in the large $\mu$,
so-called Thomas-Fermi, limit.
Additionally, the dependence of the precession frequency
on the radius of an initially displaced from the center vortex
is examined and the corresponding predictions are tested against
numerical results.
\end{abstract}

\maketitle

\section{Introduction}

In the past two decades, the study of the dynamics of
quantized vortices, few-vortex clusters and large scale vortex
lattices has seen considerable development 
due to the experimental capabilities rendered 
available in the context of atomic Bose-Einstein condensates 
(BECs)~\cite{becbook1,becbook2,siambook}.
More specifically, some of the relevant developments have
encompassed (but have not been limited to) the study of the excitation
and also the precession of few  vortices~\cite{And2000.PRL85.2857,Hal2001.PRL86.2922,Bre2003.PRL90.100403,NeelyEtAl10,dshall,dshall1,dshall3},
the observation of the instability and decay of higher charged vortices
into singly-charged ones~\cite{S2Ket2,Iso2007.PRL99.200403}, as well as
the formation of vortices and vortex rings via the transverse
instability of dark solitons~\cite{And2001.PRL86.2926,Dut2001.Sci293.663,zwierlein_ring}. At the level of large numbers of vortices, some of the
focal points have been the internal modes (collective excitations)
of vortex lattices~\cite{Eng2002.PRL89.100403,Cod2003.PRL91.100402,Smi2004.PRL93.080406,Sch2004.PRL93.210403,Tun2006.PRL97.240402}, and the study of quantum turbulence and associated
energy cascades~\cite{Hen2009.PRL103.045301,Nee2013.PRL111.235301,shin14}.
In recent works, finite temperature effects are 
also starting to be more systematically investigated, including the
formation of thermally activated vortex pairs~\cite{shin13}
and the relevant dissipation induced 
dynamics of such pairs~\cite{shin15}. Admittedly,
it is difficult to fit all the developments (even just the experimental ones)
in this partial list, yet we believe that this list provides
a substantial flavor of the wide range of relevant activity.

At the same time, the recent years have seen a tremendous increase
in the control over the experimental settings. On the one hand,
recent techniques have made it possible to ``paint'' arbitrary
types of potentials in atomic BECs~\cite{boshier}. On the other hand,
there has been a significant array of developments enabling the
extreme tunability of interactions in atomic gases via the
utilization of mechanisms such as the Feshbach resonance~\cite{hul1}.
The latter also continues to be a source of significant insights
regarding the formation of coherent structures, their
interactions, relative phases, and so on~\cite{hul2}.

The present contribution is at the interface between the two
above themes. The ability of numerous experimental groups
to construct a variety of potentials, including toroidal
ones (see Refs.~\cite{gretchen,boshier2} for some select examples), 
renders natural the question of the motion of the vortices
and their precession frequency in more general such potentials.
Our aim in the present work is to use the general methodology
of Ref.~\cite{kimfetter} (see also Refs.~\cite{lundh,svidz}) in order
to extract the equation of motion of vortices, in principle,
in arbitrary radial potentials $V(r)$. 
To obtain more concrete results, we subsequently constrain the methodology
a bit further to radial power potentials, of the general form 
$V(r)=k_p r^p$. For such potentials, we 
derive the equation of motion of a vortex, 
and restricting considerations to the vicinity of $r \rightarrow 0$,
we infer the precession frequency 
in the vicinity of the trap center. 
It is well known that in the case of the parabolic potential, 
$p=2$, this frequency has a logarithmic correction,  
namely $\omega \sim \Omega^2/(2 \mu) \ln(\mu/\Omega)$, where
$\mu$ is the chemical potential (i.e., the background density at the trap center)
and $\Omega$ sets the trap strength according to $k_2=\frac{1}{2}\Omega^2$; 
see, e.g., Ref.~\cite{svidz} and the more recent discussions of 
Refs.~\cite{middel,pelin}.
Yet, here, we find the somewhat surprising result that for general $p>2$,
and large $\mu$, the frequency decays as $\omega \sim \mu^{-2/p}$, i.e.,
there is no logarithmic correction.
We identify the general precession frequency
and corroborate numerically both
the case of $p=2$, as well as those of $p=4$ and $p=6$.
Furthermore, we explore how this precession frequency
in the immediate vicinity of the origin is modified for
a vortex located off-center and compare
these results with direct numerical simulations.

Our presentation is structured as follows. In Sec.~\ref{sec:model}, we
provide the theoretical formulation 
and the associated analytical results. In Sec.~\ref{sec:results}, 
we compare these findings with numerical results for both the stability and the dynamics. 
Finally, in Sec.~\ref{sec:conclu} we summarize our findings 
and present some challenges for future work.

\section{Model and Theoretical Analysis}
\label{sec:model}

\subsection{The Gross-Pitaevskii equation}

In the framework of mean-field theory, and for sufficiently
low-temperatures, the dynamics of a quasi-2D repulsive BECs, 
confined by a time-independent trap $V$, is described by the following
dimensionless Gross-Pitaevskii equation (GPE) (see Ref.~\cite{siambook}
for relevant reductions to dimensionless units)
\begin{eqnarray}
i \frac{\partial \psi}{\partial t}= -\frac{1}{2} \nabla^2 \psi+V \psi +| \psi |^2 \psi,
\label{eq:GPE}
\end{eqnarray}
where $\psi(x,y,t)$ is the macroscopic wavefunction.
As indicated above, the aim of our analysis will be to explore
a general radial potential $V(r)$, although we more specifically
have in mind (considering a Taylor expansion of the general
potential) a power law of the form:
\begin{equation}
\notag
V_p(r)= k_p {r}^p,
\end{equation}
where $r=\sqrt{x^2+y^2}$ denotes the radial variable. Some  of
the focal point examples in what follows (especially in our connection
with numerical computations) will consist of the cases
$p=2$, 4 and 6.
Note that the potential has rotational symmetry with respect 
to the origin, and the case of $p=2$ corresponds to the usual harmonic 
trap~\cite{becbook1,becbook2}.

In this system, we seek stationary states of the form:
\begin{eqnarray}
\psi(\vec{r},t) = \psi^{(0)}(\vec{r})e^{-i\mu t},
\notag
\end{eqnarray}
where $\mu$ is the chemical potential; substitution in Eq.~(\ref{eq:GPE})
leads to the time-independent GPE:
\begin{eqnarray}
\label{SS1}
-\frac{1}{2} \nabla^2 \psi^{(0)}+V \psi^{(0)} +| \psi^{(0)} |^2 \psi^{(0)} &=& \mu \psi^{(0)}.
\end{eqnarray}
We will seek such states
in the form of a vortex i.e., states bearing a radial profile
with $\psi^{(0)} \rightarrow 0$ as $r \rightarrow 0$ (according to
a power law $\psi^{(0)} \sim r$) and also decaying to
$0$ due to the trap effect at large $r$. The phase profile
involves a rotation by $2 \pi$ (due to their robust stability
we focus on single charge vortices) and lends the wavefunction
a structure $\psi \sim \exp(i \theta)$, where $\theta$ is the polar
variable.

\subsection{Theoretical analysis of precession frequencies}

In the work of Ref.~\cite{kimfetter} (as well as in the earlier 
one of Ref.~\cite{lundh}, and also in the review of Ref.~\cite{svidz}),
it was realized that in describing the precessing motion
of the vortex in the prototypical parabolic trap, 
it suffices to consider the vortex as bearing solely
a phase structure without considering in detail its density
profile. However, it should be mentioned in passing here
that the latter is also possible (yet it leads to rather
comparable results), as was developed using a hyperbolic
tangent approximation of the density~\cite{castin}.
In that light, we will utilize the relevant simplified
ansatz for a singly-charged (point) vortex located at
position $(x_1,y_1)$,
\begin{eqnarray}
  \psi=\psi_{\rm{TF}} \exp(i S), \quad S=\tan^{-1} \left(\frac{y-y_1}{x-x_1} \right),
  \label{eqn1}
\end{eqnarray}
where $\psi_{\rm{TF}}=\sqrt{\max(\mu-V,0)}$ is the Thomas-Fermi (TF) ground state. By means of this variational approximation for the wavefunction, it is possible to identify the kinetic energy of the vortex field as~\cite{lundh}:
\begin{eqnarray}
  T&=&\frac{i}{2} \int_{\mathbb{R}^2} (\psi^{*} \psi_t - \psi_t^{*} \psi) \, dx \,dy,
\notag
\\[1.0ex]
\notag
  &\approx& -2 \pi \dot{\phi_1} \int_0^{r_1} [\mu - V(a)] a\, da,
\end{eqnarray}
where $r_1=\sqrt{x_1^2+y_1^2}$ is the distance of the vortex
to the center of the trap and $\phi_1$ is the polar angle for the
vortex position (star denotes complex conjugate and overdot 
differentiation with respect to $t$).
Using the ansatz (\ref{eqn1}), it
is also possible to express the (potential) 
energy of the system:
\begin{eqnarray}
  E &=& \int_{\mathbb{R}^2}  \left[\frac{1}{2} |\nabla \psi|^2 + V |\psi|^2 +
  \frac{1}{2} |\psi|^4 \right] dx \, dy
  \nonumber \\
  &\approx& \pi \left[\int_0^{r_1-\xi} r\,\frac{\mu-V(r)}{r_1^2-r^2} dr
+  \int_{r_1+\xi}^{R_{\rm TF}} r\, \frac{\mu-V(r)}{r^2-r_1^2} dr\right],
  \nonumber \\
  \label{poten} 
\end{eqnarray}
where $\xi=1/\sqrt{2\mu}$ is vortex core width (given by the healing
length) and $R_{\rm TF}$ is the TF radius such that $V(R_{\rm TF})=\mu$.
As proposed in Ref.~\cite{kimfetter}, the above result is
based on the regularization of the energy integral by removing
a ``layer'' of width $2\xi$ about the singularity at $r_1$.
In the present work we extend this methodology
to general radially symmetric potentials.
Details of the relevant derivation are
provided in Appendix \ref{appendix1}. 

Considering now the Lagrangian $L=T-E$, one can obtain the
resulting dynamical equation of motion for the vortex
precession. In this case, the obtained evolution is along the
azimuthal direction and of the form~\cite{lundh}:
\begin{eqnarray}
  \dot{\phi}_1= \frac{-1}{2 \pi r_1 \left( \mu - V(r_1) \right)}
  \frac{\partial E}{\partial r_1}.
  \label{dynam}
\end{eqnarray}
While it is clear that the radial potential will generically
result in precessional dynamics, it is instructive to consider
some special case examples (including the well known, experimentally
relevant one of the parabolic trap as a benchmark).

\subsection{Parabolic Trap $V(r)=\frac{1}{2}\Omega^2 r^2$}

For a parabolic trap $V(r)=\frac{1}{2}\Omega^2 r^2$, the energy reads:
\begin{eqnarray}
  E&=& \frac{\pi \Omega^2}{4} \left[2 (\xi^2 + r_1^2) - R_{\rm TF}^2 \phantom{\frac{X}{X}} \right.
\notag
\\[1.0ex]
\notag
&+& \left. (r_1^2 -R_{\rm TF}^2) \ln \left( \frac{\xi^2 (4 r_1^2-\xi^2)}{r_1^2 ( R_{\rm TF}^2 - r_1^2)} \right)  \right],
\end{eqnarray}
where $R_{\rm TF}^2=2 \mu/\Omega^2$. The above expression for the energy 
leads, via Eq.~(\ref{dynam}), to the following equation of motion:
\begin{eqnarray}
\notag
  \dot{\phi}_1=\frac{4 r_1^2-\xi^2 R_{\rm TF}^2 - r_1^2 (\xi^2-4 r_1^2)
    \ln\frac{\xi^2(4 r_1^2-\xi^2)}{r_1^2 (R_{\rm TF}^2 - r_1^2)}}{2 r_1^2
    (\xi^2 -4 r_1^2) (R_{\rm TF}^2 - r_1^2)}. 
\end{eqnarray}
If we now consider the motion near the center, taking large
chemical potential so that $1/\sqrt{2 \mu} \equiv \xi \rightarrow 0$,
yet also $r_1 \rightarrow 0$ (with $r_1 \gg \xi$), 
we retrieve the well-known result~\cite{svidz}
according to which the precession frequency is approximated as:
\begin{eqnarray}
  \omega= \frac{\Omega^2}{2 \mu} \ln\left(\frac{\mu}{\Omega}\right).
  \label{parab2}
  \end{eqnarray}
Notice that subsequent works (see, e.g., Ref.~\cite{middel}) devised
numerically inspired corrections to this formula ---although
not to its functional form---, yet
it has been particularly successful in capturing the functional
form of the dependence on the chemical potential $\mu$ (and the
frequency $\Omega$).

\subsection{The Quartic Potential $V(r)=k_4 r^4$}

For a quartic potential, $V(r)=k_4 r^4$, the TF radius is given by
$k_4 R_{\rm TF}^4=\mu$. In this case, 
the potential energy can still be calculated analytically:  
\begin{eqnarray}
  E &=& \frac{\pi k_4}{4}
  \left[ 2 \xi^4+ 16 \xi^2 r_1^2 +6 r_1^4 -2 r_1^2 R_{\rm TF}^2-R_{\rm TF}^4
     \phantom{\frac{X}{X}} \right.
    \nonumber
    \\
    \nonumber
    &+& \left. 2 (r_1^4-R_{\rm TF}^4) \ln \left(\frac{\xi^2(4 r_1^2-\xi^2)}{r_1^2(R_{\rm TF}^2-r_1^2)}\right) \right].
\end{eqnarray}
Naturally, the dynamics can be extracted from Eq.~(\ref{dynam}),
yet it is too unwieldy and not particularly informative to provide here.
Instead, we focus once again on the limit of
$\xi,r_1 \rightarrow 0$, with $\xi$ tending faster to the limit.
The remarkable observation here, and in general for other
powers $p>2$ that we have examined, the logarithmic term tends to
$0$ (due to its proportionality to some power of $r_1$). Hence, 
the logarithmic correction {\it does not survive} as it does 
in the parabolic case. Instead, in this case, the limit reads:
\begin{eqnarray}
  \dot{\phi}_1 =\omega = \frac{1}{R_{\rm TF}^2} \sim 
%\sqrt{k_4}\, 
{\mu}^{-\frac{1}{2}}.
  \label{quartfreq}
\end{eqnarray}

\subsection{General Power $V(r)=k_p r^p$}
\label{sec:sub:gral}

For a radially symmetric potential $V(r)=k_p r^p$ with general power $p$,
we have $\mu=k_p R_{\rm TF}^p$. Remarkably, the general
form of the energy is again available in analytic form for arbitrary $p$:
\begin{widetext}
\begin{eqnarray}
  E &=& \frac{\pi k_p}{2}
    \left[
     r_1^p \left(B\left(\frac{r_1^2}{R_{\rm TF}^2},-\frac{p}{2},0\right)-
     B\left(\frac{r_1^2}{(\xi+r_1)^2},-\frac{p}{2},0\right)\right)
    \right.
    \nonumber
    \\
    \label{generalen}
    &-& 
    \left.
         2 (r_1-\xi)^{2+p}
        \,{_2}F_1\left(1,\frac{p+2}{2},\frac{p+4}{2},\left(\frac{r_1-\xi}{r_1}\right)^2\right)
%    \nonumber
%    \\
%    &+& \left. 
% + R_{\rm TF}^p \ln \left( \frac{r_1^2 (r_1^2-R_{\rm TF}^2)}{\xi^2(\xi^2-4 r_1^2)}\right)
 - R_{\rm TF}^p \ln \left( \frac{\xi^2 (4 r_1^2-\xi^2)}{r_1^2(R_{\rm TF}^2-r_1^2)}\right)
   \right],
    %\right]
\end{eqnarray}
\end{widetext}
where $B$ denotes the incomplete Beta function, while $F$ denotes
the hypergeometric function.
By considering integer $p$, the resulting asymptotics
in the limit of $\xi \rightarrow 0$, and $r_1 \rightarrow 0$
(with $r_1 \gg \xi$)
from the gradient of $E$ [in Eq.~(\ref{generalen})] leads to
\begin{eqnarray}
  \dot{\phi}_1 = \omega = \frac{p}{2(p-2)} \frac{1}{R_{\rm TF}^2} \sim \mu^{-\frac{2}{p}},
  \label{final}
\end{eqnarray}
which is consonant with the $p=4$ result of Eq.~(\ref{quartfreq}).
In the general asymptotic form of Eq.~(\ref{generalen}), the corresponding
prefactor $ \frac{p}{2(p-2)}$
is evidently different for different values of $p$,
but, importantly, the scaling relation is general providing an explicit
power law prediction for the dependence of the precession frequency
on the chemical potential (i.e., the background density which
also controls the width/healing length scale of the vortex).

\begin{figure}[tb]
\begin{center}
\includegraphics[width=8.5cm]{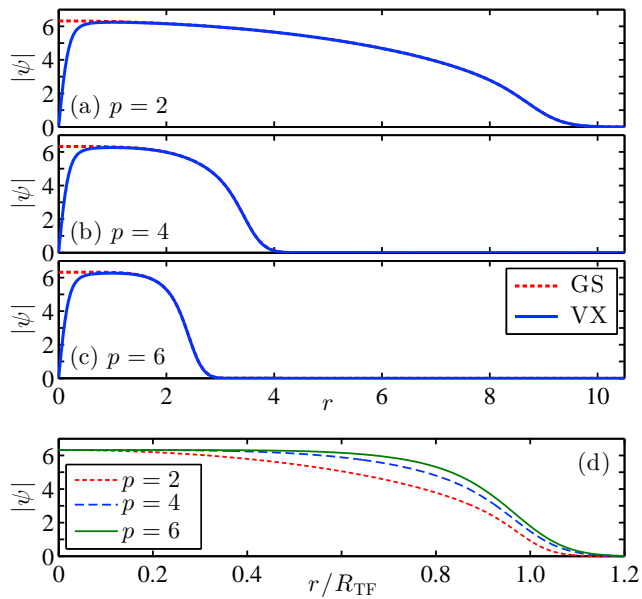}
\caption{
(Color online)
Radial profile $|\psi|$ as a function of $r$ for the ground state (GS, red dashed line) 
and the vortex state (VX, blue solid line) inside the trapping potential 
$V(r)=\frac{1}{p}r^p$ with $\mu=40$ for (a) $p=2$, (b) $p=4$, and (c) $p=6$. 
Note that in the Thomas-Fermi limit, the size of the vortices for 
the different potentials is essentially the same. This property, determined
by the healing length, is controlled by the chemical potential $\mu$
which is constant ($\mu=40$) for the three cases.
Panel (d) depicts the GS profiles as a function of the rescaled
distance $r/R_{\rm TF}$ where the corresponding TF radii are
given by $R_{\rm TF}=(p\mu)^{1/p}$.
}
\label{States}
\end{center}
\end{figure}

It is important to stress that the logarithmic term in the energy, and hence 
in the precession frequency, is always present independently of
the chosen power $p$ of the confining potential. This logarithmic term is 
crucial (dominant) for parabolic trapping potentials (see Eq.~(\ref{parab2}) 
and Refs.~\cite{svidz,Fetter_review_2009}).
However, as shown above, for potentials with powers larger than quadratic,
the logarithmic term decays faster than the remaining terms in the
asymptotic expression near the origin, and hence it is no longer
dominating the relevant asymptotics.
Nonetheless, it should be pointed out that far from the origin, the
logarithmic correction terms will indeed become significant and the full 
expression without discarding these terms needs to be used.

In order to complement the above result for the precession frequency using
the asymptotics for the energy, we have also employed a direct matching asymptotics
analysis on Eq.~(\ref{eq:GPE}) that yields precisely the same
asymptotic result as in Eq.~(\ref{final}).
Details on this matched asymptotic procedure can be found in
Appendix \ref{appendix2}.
We now turn to a numerical examination of the relevant findings.

\section{Numerical Results}
\label{sec:results}

In our numerical work, we study $V=\frac{1}{p}r^p$, where $p=2, 4$ and 6. 
We start by showing the ground and vortex states at a typical (relatively large)
value of the chemical potential, $\mu=40$, within the so-called
Thomas-Fermi regime. In this regime, the radial background density 
(i.e., the density of the ground state) can be well approximated as
$|\psi|^2 \approx |\psi_{\rm TF}|^2 = \mu - V(r)$. Relevant results as depicted
in Fig.~\ref{States}. One can observe that the states near $r=0$ are almost 
identical for different potentials. In particular, the density of the ground 
states at $r=0$ and the width of the vortices is essentially dominated by 
the chemical potential which controls the corresponding healing length.
The size of the states gets smaller as $p$ increases or, equivalently,
as the atoms are bound tighter.

\begin{figure}[tb]
\begin{center}
\includegraphics[width=8.5cm]{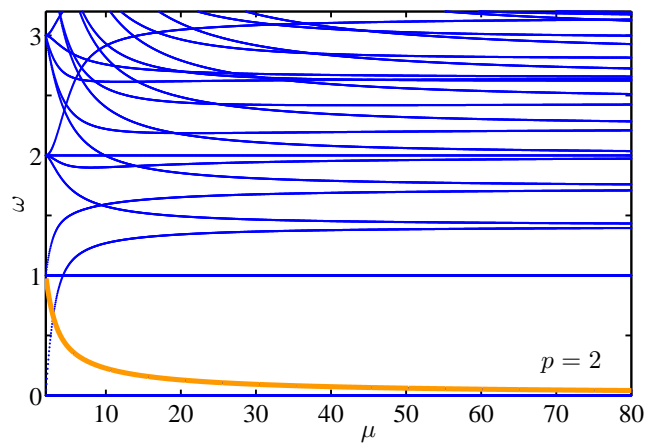}
\caption{
(Color online)
  The BdG spectrum for the vortex in the quadratic potential.
  The lowest mode, highlighted in orange, is the one of 
  interest as it corresponds to
  the vortex precession around the trap center.
}
\label{V2}
\end{center}
\end{figure}

\begin{figure}[tb]
\begin{center}
\includegraphics[width=8.5cm]{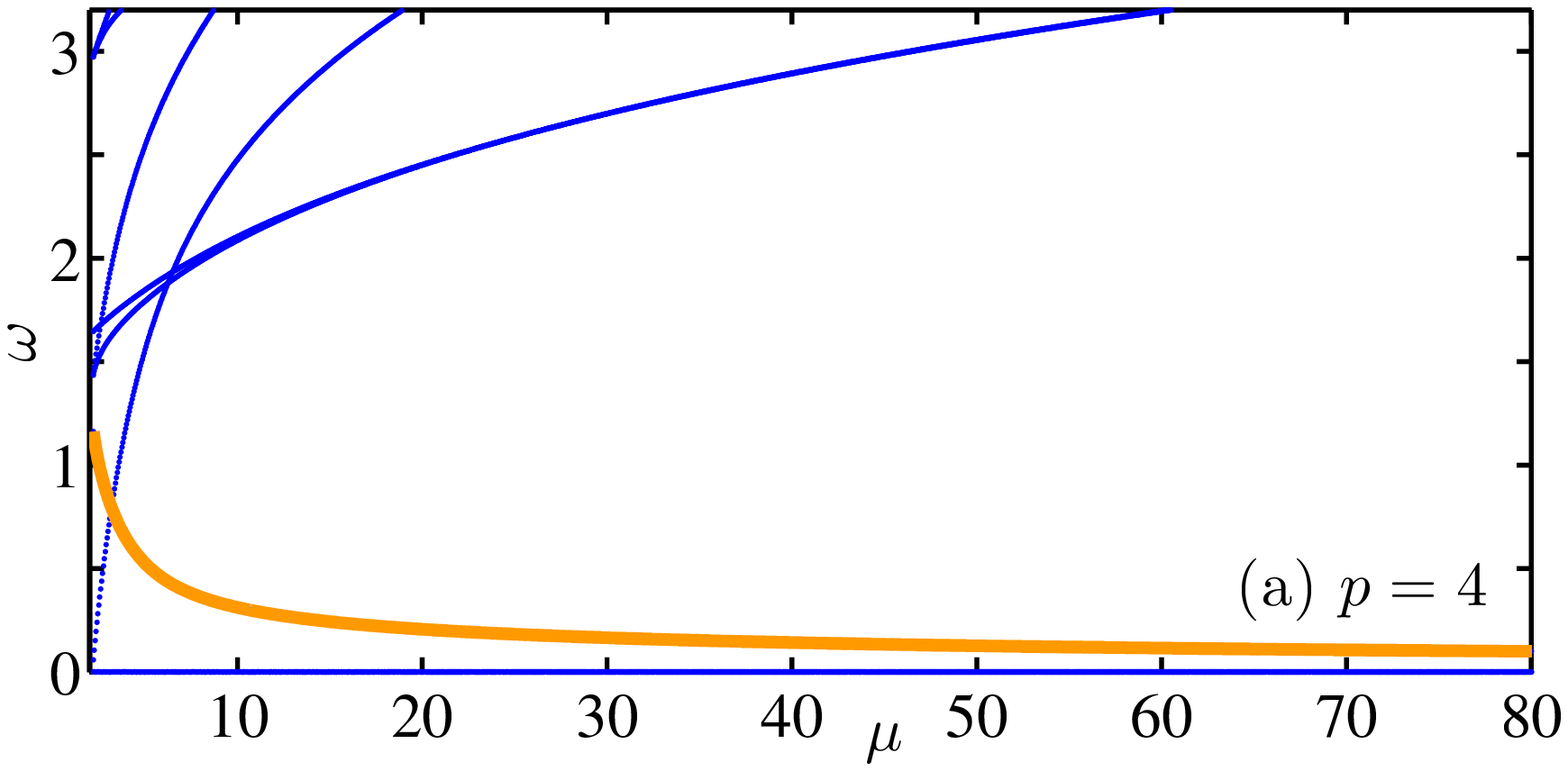}
\includegraphics[width=8.5cm]{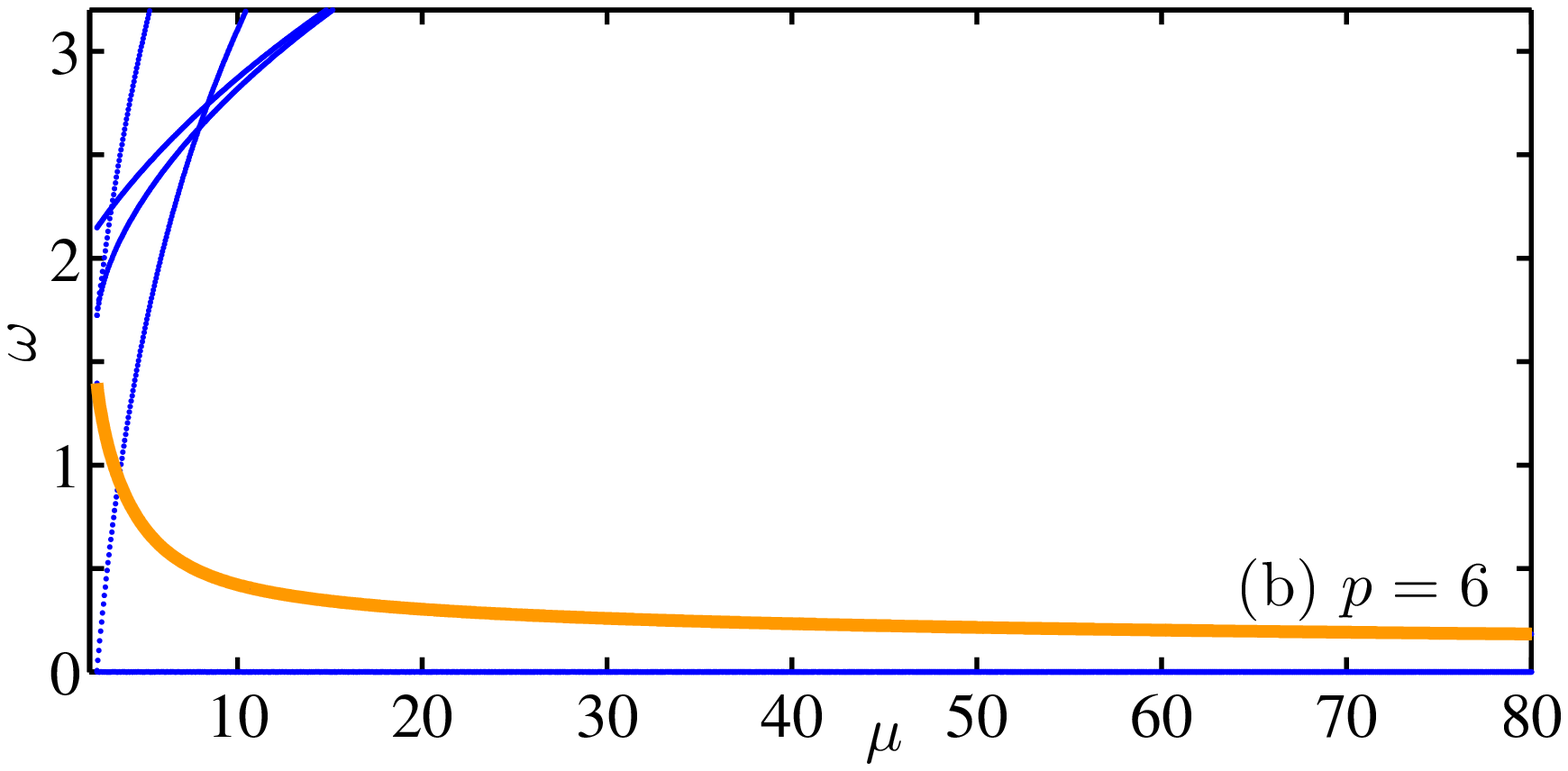}
\caption{
(Color online)
  The vortex spectrum in the (a) quartic and (b) sextic potentials.
  The lowest mode which is highlighted in orange similar to the quadratic potential is the vortex precession mode around the trap center.
}
\label{V4}
\end{center}
\end{figure}

%\subsection{The parabolic potential}

Let us start by showing the numerical results for the parabolic potential
$V(r)=\frac{1}{2}r^2$ (i.e., $\Omega=1$).
The Bogolyubov-de Gennes (BdG) stability spectrum for the steady state 
consisting of a unit-charge vortex at the center of the trap
is depicted in Fig.~\ref{V2}. Among all the modes in the spectrum,
the lowest is the one that is {\it not} found in the
spectrum of linearization around the ground state.
Instead, when we excite this mode, we find that it leads the vortex to
a precessional motion around the center of the trap; 
thus, the frequency
of this mode corresponds to the frequency of the associated
precession~\cite{middel}.
This precessional mode is depicted with a orange line in Fig.~\ref{V2}.

%\subsection{The quartic and sextic potential}

We now go beyond the well-known parabolic case, in order to
examine other case examples that the general theory can tackle.
More specifically, we consider the quartic case of $p=4$ and
the sextic case of $p=6$.
The BdG spectra for the corresponding steady state vortex configurations are shown
in Fig.~\ref{V4}. Note that both potentials still have the vortex precession 
mode (see eigenfrequency mode depicted in orange), because of the symmetry 
of the potential.
Indeed, we find that in both cases it remains the only mode asymptoting
to $\omega=0$, as the chemical potential $\mu$ is increased.
The case examples of the associated dynamics that we have
chosen to show are for $\mu=40$. This is much larger than that of 
the case $p=2$, because the cases $p=4$ and $p=6$ involve a tighter binding trap. 
It is interesting to note that the vortex in all three cases is remarkably 
stable for all the values of the chemical potential that we have examined.

\begin{figure}[tb]
\begin{center}
\includegraphics[width=8.5cm]{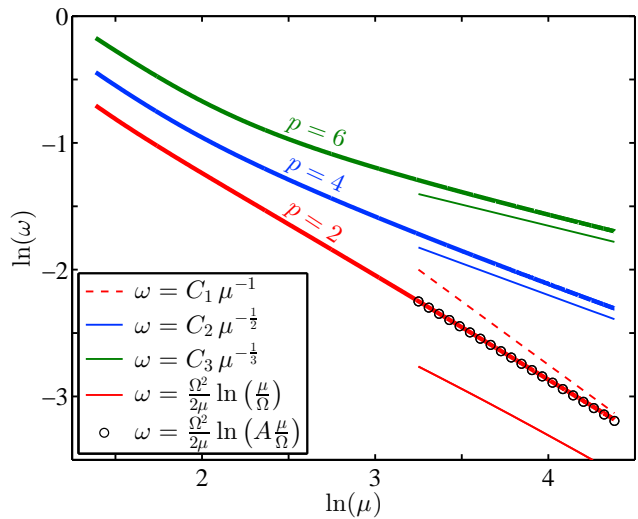}
\caption{
(Color online)
  Scaling of the precession frequency $\omega$ on the chemical
  potential $\mu$. The thick lines corresponds to the precession frequency
  extracted from the numerically obtained spectra and the asymptotic approximations
  are depicted with thin lines ($C_i$ are arbitrary constants chosen for
  ease of exposition).
  The case of $p=2$ decays slower compared with the
  power scaling (see thin red dashed line) due to the logarithmic correction, while
  for $p=4$ and $p=6$, the predictions and the spectrum appears to compare well 
  with each other, suggesting a good agreement with the analytical
  prediction of Eq.~(\ref{final}).
  %
  %The thin black line corresponds to the logarithmic asymptotic formula (\ref{parab2}).
  The black circles correspond to a correction to the asymptotic formula (\ref{parab2}),
  namely Eq.~(\ref{eq:parab_correct}) with $A=8.88$, see Ref.~\cite{middel}.
  }
\label{NV}
\end{center}
\end{figure}

%\subsection{Frequency Dependence on Chemical Potential}

Let us now focus on the scaling of 
the precession frequency $\omega$ (close to the center of the trap) on
the chemical potential for all three cases of $p=2$, $4$ and $6$. 
The corresponding results are depicted in Fig.~\ref{NV}
and are compared to our theoretical prediction,
as encapsulated in Eq.~(\ref{final}) %
\footnote{It is worth
mentioning that a key feature of our numerical results for the computation of the BdG
spectra is that we use numerical
methods that we have described in earlier works~\cite{shell},
involving a quasi-one-dimensional radial computation (for different azimuthal
wavenumbers). This allows us to explore large values of
the chemical potential so as to reach the TF
limit, where our analytical prediction is relevant
(since there the internal density structure of the vortex can be ignored).}.
In this figure, the relevance of our analytical prediction, and especially
of the corresponding scaling is evident.
Indeed, once the (small and) intermediate chemical potential regime (where 
all potentials scale similarly) is bypassed and the large chemical potential 
TF regime is reached, the different potentials scale differently.
More specifically, it is found that the scaling prediction of Eq.~(\ref{final})
is closely followed by the spectral numerical results for $p=4$ and $p=6$
(see, respectively, the thin solid blue and green lines). 
On the other hand, for $p=2$, the prediction of Eq.~(\ref{final})
(see red dashed line) is incorrect as the dominant term in this 
case has a logarithmic form, see Eq.~(\ref{parab2}) and the thin solid 
red line in Fig.~\ref{NV}.
It is worth mentioning that, although the precession frequency predicted
by Eq.~(\ref{parab2}) has the correct scaling, a numerical 
factor has been used in order to incorporate
the sub-dominant contributions to the frequency scaling
according to $\mu^{-1}$~\cite{middel} in a quantitative fashion.
This is depicted by the black circles in the figure corresponding to
\begin{equation}
\label{eq:parab_correct}
\omega = \frac{\Omega^2}{2 \mu} \ln\left(A\frac{\mu}{\Omega}\right), 
\end{equation}
with the numerical factor $A=8.88$, see Ref.~\cite{middel}.

\begin{figure}[tb]
\begin{center}
\includegraphics[width=8.5cm]{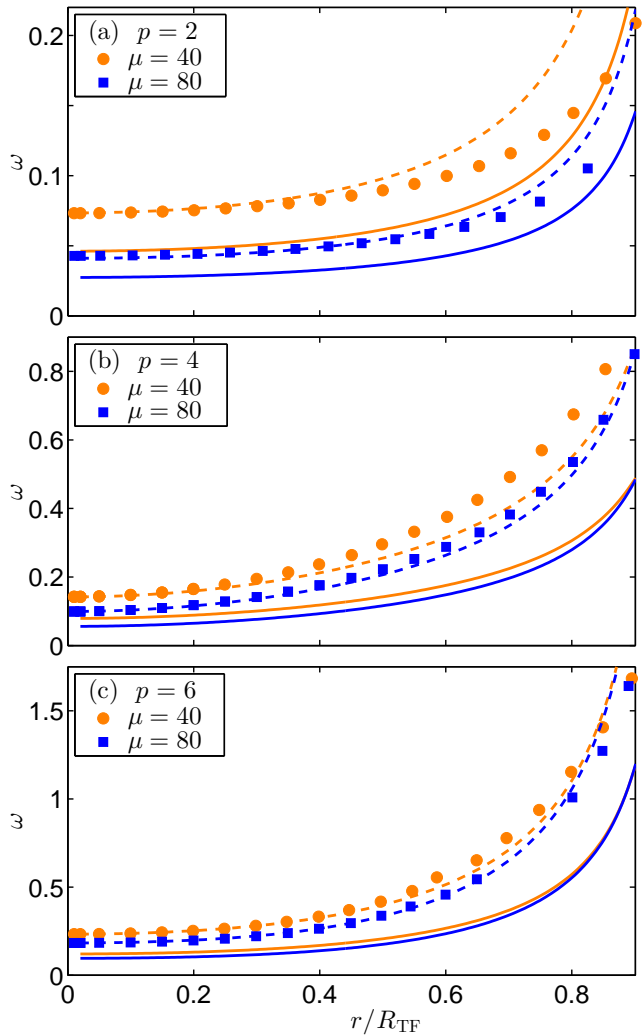}
\caption{
(Color online)
Scaling of the precession frequency as a function of the rescaled vortex 
position for the different potentials $V(r)=\frac{1}{p} r^p$ with 
$\mu=40$ and $\mu=80$. 
A vortex is displaced from $r=0$ and its precession frequency is extracted 
from its dynamical evolution (see text) and depicted with orange (blue) 
circles (squares) for $\mu=40$ ($\mu=80$). 
The solid lines depict the theoretical predictions 
from evaluating Eq.~(\ref{dynam}) and using the limit $\xi\ll r$.
The dashed curves for the $p=2$ case correspond to the corrected
precession of Eq.~(\ref{eq:parab_correct}) while the dashed curves
for the $p=4$ and $p=6$ cases
depict the rescaled theoretical predictions
so  as to match the precession frequency in the limit $r=0$.
These rescaling coefficients correspond to,
respectively for $\mu=40$ and $\mu=80$,
%%factors_p2 = 1.8157    1.7045    1.6041
%%factors_p4 = 1.7944    1.7804    1.7669
%%factors_p6 = 1.9302    1.9114    1.9035
%$p=2$: 1.8157 and 1.7045,
$p=4$: 1.7944 and 1.7804, and
$p=6$: 1.9302 and 1.9114.
}
\label{EP}
\end{center}
\end{figure}

To complement the description for the vortex precession around the 
center of the trap, we also measure the dependence of this frequency
as the vortex is shifted away from $r=0$.
In practice, a straightforward way to observe this precession, involves
shifting the vortex off of its equilibrium position at $r=0$
and then following its circular motion around the center.%
\footnote{We should mention that
the actual vortex orbit is not exactly circular as the vortex pushes
out a small amount mass away from its core and, as it precesses, 
dipolar and quadrupolar modes of the background cloud are weakly excited. 
Nevertheless, it is possible to accurately measure the 
oscillation frequency by following the dynamics for a sufficiently 
long time, here we typically integrate all dynamics up to $t=1000$,
and applying a least-square fit using a sine function.
We identify the location of the vortex during the dynamics by 
looking for the density minimum in the neighborhood of the vortex 
using a finer grid cubic spline interpolation.}
Figure~\ref{EP} depicts the departure of the precession frequency 
from the corresponding eigenfrequency, $\omega_0$, at $r=0$  
measured from the center of the trap.
The figure also depicts the analytical prediction given 
from evaluating Eq.~(\ref{dynam}) and using the limit $\xi\ll r$
(see solid curves in the figure).
It is evident from these results that the theory underestimates the precession 
frequency. We attribute this discrepancy to two possible factors:
\begin{itemize}
\item[(a)]
The ansatz used in the theory completely
disregards the internal spatial
structure of the vortex as it effectively treats it as a point vortex,
an approximation valid as $\mu \rightarrow \infty$.
For instance, optimizing the vortex width
appears to be an important factor in the success of
more refined (yet less straightforwardly tractable in the general
case) variational approaches such as that of Ref.~\cite{castin}.
\item[(b)]
Secondly, when estimating the precession frequency at the center of 
the trap ($r=0$) the limit $0 < \xi \ll r$ is obviously violated as 
the healing length $\xi$ is finite.
\end{itemize}
Although the theoretical results fail to precisely predict
the precession frequency at $r=0$, they are able to give the
right tendency for the departure of the precession frequency 
as the vortex is displaced away from the origin.
In fact, after applying suitable modifications (see dashed
curves in Fig.~\ref{EP}), the predictions produce a good quantitative 
match for the departure of the precession frequency from
$\omega_0$ as $r$ departs from $r=0$ (especially for small 
and intermediate values of $r$ and progressively better
for larger values of $\mu$, in line with the underlying premise
of the theory).
Specifically, in the $p=2$ case, where the dominant term on the precession
as a function of $r$ is logarithmic, we employ the correction
given in Eq.~(\ref{eq:parab_correct}) Ref.~\cite{middel}.
On the other hand, for the $p=4$ and $p=6$ cases, the dominant
terms are algebraic and thus we opt for a multiplicative
rescaling factor chosen so as to match the precession frequency
at the origin (see Fig.~\ref{EP}).

\section{Conclusions and Future Challenges}
\label{sec:conclu}

In the present work, we have explored the motion of vortices
in general radially symmetric potentials. We have found that similarly
to the well known parabolic case, the motion of the vortices
involves a precession. The main result of the present work
concerns the vortex precession frequency and its dependence on both
the chemical potential of the background cloud and the location
of the vortex within the condensate. Utilizing a variational
formulation generalizing the work of Ref.~\cite{svidz},
we were able to provide closed form expressions
for the precession frequency relevant in the large chemical potential
limit (Thomas-Fermi regime). These expressions permitted us to
appreciate the power law scaling of the precession frequency on the
chemical potential and how the relevant dependence becomes slower
as $p$ increases. The limitations of the theory in identifying
the precession mode frequency at the origin were explained
and it was shown how a suitable amendment can be used to capture
its dependence on the radial position of an off-center vortex.

While in the present work we have explored the general (radial) potential
motion of a single vortex, numerous related questions naturally
emerge from this study. Here, we considered isotropic potentials
which still constitute a subject of active
investigation~\cite{Esposito}
and some controversy regarding the physical interpretation
of the origin of the vortex motion~\cite{simula}.
Yet, anisotropic potentials are
also quite relevant in atomic BECs~\cite{svidz,siambook}.
Examining vortex motion in such anisotropic settings under
general $V$ would be certainly of interest. Furthermore, combining
an understanding of the single vortex motion in a general
potential with that of the inter-vortex interaction will enable
identifying multi-vortex (cluster and crystal) states for
arbitrary trapped BEC systems. 
Another possibility for future research could be to consider
pseudo-potentials stemming from considering space (radial) dependent
nonlinearities \cite{Boris_pseudo_potential} with different
power law prescriptions.

Finally, while the above ideas
are natural to be first developed in two-dimensional settings, generalizing
them to vortex rings in three-dimensional frameworks~\cite{komineas}
would also be of interest in its own right. Some of these directions
are currently under consideration and will be reported in future
publications.

\begin{acknowledgments}

We thank A. Esposito, R. Krichevsky, and A. Nicolis for bringing
up their related work on vortex precession in trapped superfluids 
from effective field theory~\cite{Esposito}
and for subsequent stimulating discussions.

W.W.~acknowledges support from NSF-DMR-1151387.
P.G.K.~gratefully acknowledges the support of
NSF-DMS-1312856, NSF-PHY-1602994, as well as from
the ERC under FP7, Marie
Curie Actions, People, International Research Staff
Exchange Scheme (IRSES-605096) and the Greek Diaspora
Fellowship Program. P.G.K.~also acknowledges useful
discussions with Prof.~T.~Kolokolnikov.
R.C.G.~acknowledges support from NSF-DMS-1309035 and PHY-1603058.
The work of W.W. is supported in part by the Office of the Director of National Intelligence (ODNI), Intelligence Advanced Research Projects Activity (IARPA), via MIT Lincoln Laboratory Air Force Contract No.~FA8721-05-C-0002. The views and conclusions contained herein are those of the authors and should not be interpreted as necessarily representing the official policies or endorsements, either expressed or implied, of ODNI, IARPA, or the U.S.~Government. The U.S.~Government is authorized to reproduce and distribute reprints for Governmental purpose notwithstanding any copyright annotation thereon. We thank Texas A\&M University for access to their Ada and Curie clusters.

\end{acknowledgments}

%%%%%%%%%%%%%%%%%%%%%%%%%%%%%%%%%%%%%%%%%%%%%%%%%%%%%%%%%%%%%%%%%%%%%%%%%%%%%

\appendix

\section{Energy calculation}
\label{appendix1}

As explained in Ref.~\cite{kimfetter}, the principal
contribution to the energy stems from the gradient term,
which upon our ansatz (\ref{eqn1}) and in the TF approximation
can be written as:
\begin{widetext}
\begin{eqnarray}
\notag
  E= \frac{\mu}{2} \int_0^{R_{\rm TF}}  r dr \int_0^{2 \pi} d\theta
  \left(1- \frac{V(r)}{\mu}\right)
  \frac{1}{r^2 + r_1^2 -2 r r_1 \cos(\theta-\phi_1)},
\end{eqnarray}
where, again, $(r1,\phi_1)$ is the vortex position in polar coordinates.
This integral can be split into two radial contributions
%(the azimuthal part yields a factor of $2 \pi$ for
%a radially symmetric potential in both),
namely the integral $I_1$ from $0$ to $r_1-\xi$
and $I_2$ from $r_1+\xi$ to $R_{\rm TF}$. This way, using
the characteristic length of the vortex core, namely the healing length $\xi$,
we regularize the integral. These two integrals can then be
rewritten as:
\begin{eqnarray}
  I_1&=&\frac{1}{2} \int_0^{r_1-\xi} dr \int_0^{2 \pi} d\theta\, \frac{r (\mu-V(r))}{r_1^2-r^2}
  \left(1 + 2 \sum_n \left(\frac{r}{r_1}\right)^n \cos(n (\theta-\phi_1)) \right),
\notag
\\
\notag
  I_2&=&\frac{1}{2} \int_{r_1+\xi}^{R_{\rm TF}} dr \int_0^{2 \pi} d\theta\, \frac{r (\mu-V(r))}{r^2-r_1^2}
  \left(1 + 2 \sum_n \left(\frac{r_1}{r}\right)^n \cos(n (\theta-\phi_1)) \right),
\end{eqnarray}
\end{widetext}
which finally yield the following resulting expressions 
used in Eq.~(\ref{poten}):
\begin{eqnarray}
  I_1 &=& \pi  \int_0^{r_1-\xi} dr \frac{r (\mu-V(r))}{r_1^2-r^2},
\notag
  \\
\notag
  I_2 &=& \pi  \int_{r_1+\xi}^{R_{\rm TF}} dr \frac{r (\mu-V(r))}{r^2-r_1^2}.
\end{eqnarray}

%%%%%%%%%%%%%%%%%%%%%%%%%%%%%%%%%%%%%%%%%%%%%%%%%%%%%%%%%%%%%%%%%%%%%%%%%%%%%

\section{Matched asymptotics calculation}
\label{appendix2}

We hereby obtain an alternative derivation for the precession
frequency of Eq.~(\ref{final}) using matched asymptotics directly 
on the original model~(\ref{eq:GPE}).
We seek for stationary solutions to Eq.~(\ref{eq:GPE}) of the form 
$\psi=\psi^{(0)} e^{-i\mu t}$ satisfying the steady state Eq.~(\ref{SS1}).
By scaling variables using $\tilde{x}={x}/R_{\rm TF}$, $\tilde{y}={y}/R_{\rm TF}$,
$\tilde{\psi}={\psi^{(0)}}/\sqrt{\mu}$, and $\tilde{t}={t}/(2R_{\rm TF}^2)$, we obtain:
\begin{equation}
\notag
-i\psi_t= \Delta \psi+\frac{1}{\varepsilon^2}\left(1-V(r)-|\psi|^2\right)\psi,
\end{equation}
with $\varepsilon^2={1}/(2R_{\rm TF}^2 \mu)$ and where, for ease of exposition,
we have dropped tildes and superscripts.
Note that the core size of the vortex is of order $\varepsilon$.

Away from the core of the vortex, to leading order in $\varepsilon$, 
we have $1-V(\vec{r})-|\psi(\vec{r})|^2=0$, which suggests the separation
\begin{equation}
\notag
\psi=\psi_{\rm TF}\, e^{iS}
\end{equation}
where $\psi_{\rm TF}=\sqrt{1-V(\vec{r})}$ is the TF approximation and $S$ satisfies:
\begin{eqnarray}
\psi_{\rm TF}\Delta S+2\nabla \psi_{\rm TF} \cdot \nabla S=0,
\notag
\\[1.0ex]
\notag
\nabla \times \nabla S=2\pi\delta(\vec{r}-\vec{r}_1), 
\end{eqnarray}
where $\delta$ is the Dirac-delta function centered
at the vortex location $\vec{r}_1$.
% [which, in turn, is expressed in polar coordinates as $(r_1,\phi_1)$]. 
%
By examining the local behavior of $S$, we can obtain:
\begin{widetext} %%%%%%%%%%%%%%%%%%%%%%%%%%%%%%%%%%%%%%%%%%%%%%%%%%%%%%%%%%%
\begin{equation}\label{out}
\psi(\vec{r})\sim \left(\psi_{\rm TF}(\vec{r}_1)+ \nabla \psi_{\rm TF}(\vec{r}_1){\cdot (r-\vec{r}_1)   }\right) {e^{i\varphi}}\left(1+i\ln|\vec{r}-\vec{r}_1| {\frac{\nabla^\bot \psi_{\rm TF}(\vec{r}_1)}{\psi_{\rm TF}(\vec{r}_1)} } \cdot (\vec{r}-\vec{r}_1)+\vec{K}\cdot(\vec{r}-\vec{r}_1) \right), ~~{\rm as}~~ \vec{r}\rightarrow \vec{r}_1,
\end{equation}
where $\vec{K}\equiv\lim_{\vec{r}\rightarrow \vec{r}_1} \nabla \left(S-{\varphi}\right)$,
$\varphi={\rm ang}(\vec{r}-\vec{r}_1)$, and
the operator $(\cdot)^\bot$ is defined, in Cartesian coordinates, by $(a,b)^\bot\equiv(-b,a)$.

Near the core of the vortex, we denote the stretched variable 
$\vec{\rho}=(\vec{r}-\vec{r}_1(t))/{\varepsilon}$ and look for the solution in the form:
\begin{equation}
\notag
\psi=\psi_0({\vec{\rho}})+\varepsilon\psi_1({\vec{\rho}})+\cdots
\end{equation} 
Matching the first two orders of $\varepsilon$ yields:
\begin{eqnarray}
0&=& \Delta_{{\vec{\rho}}} \psi_0+(1-V(|\vec{r_1}|))\psi_0-|\psi_0|^2 \psi_0
\notag
\\[1.0ex]
\notag
{i\,\dot{\vec{r}}_1} \cdot \nabla_{{\vec{\rho}}} \psi_0 {+\nabla V(\vec{r}_1)}\cdot {\vec{\rho}} \psi_0
&=&\Delta_{{\vec{\rho}}} \psi_1+{(1-V(|\vec{r}_1)|)} \psi_1-|\psi_0|^2 \psi_1  
-\psi_0\left(\psi_0 \psi^*_1+ \psi_1 \psi^*_0  \right),
\end{eqnarray}
where the overdot denotes time derivative.
Now, in order to match with the outer region, we only need the asymptotic behavior 
of the inner solution as $|{\vec{\rho}}| \rightarrow \infty$. 
A detailed analysis for this asymptotics yields~\cite{Shuangquan_preprint}
\begin{align*}
\psi_0 &\rightarrow \psi_{\rm TF} e^{i {\theta}}, 
&~~{\rm as}~~ |{\vec{\rho}}|\rightarrow \infty,
\\
\psi_1 &\rightarrow \left( \nabla \psi_{\rm TF}(\vec{r}_1)\cdot {\vec{\rho}}+{i}\ln(\psi_{\rm TF} |{\vec{\rho}}|) \nabla^\bot \psi_{\rm TF} \cdot {\vec{\rho}}{+}\frac{1}{2}\psi_{\rm TF}(\vec{r}_1) \dot{\vec{r}}_1 \cdot {\vec{\rho}}  \right)e^{i{\theta}}, 
&~~{\rm as}~~ |{\vec{\rho}}|\rightarrow \infty,
\end{align*}
and therefore
\begin{equation}\label{in}
\psi=\psi_{\rm TF}(\vec{r}_1) e^{i {\theta}}+\varepsilon \left( \nabla \psi_{\rm TF}(\vec{r}_1)\cdot {\vec{\rho}}+{i}\ln(\psi_{\rm TF} |{\vec{\rho}}|) \nabla^\bot \psi_{\rm TF}(\vec{r}_1) \cdot {\vec{\rho}}{+}\frac{1}{2}\psi_{\rm TF}(\vec{r}_1) \dot{\vec{r}}_1 \cdot {\vec{\rho}}  \right)e^{i{\theta}},~~{\rm as}~~|{\vec{\rho}}|\rightarrow \infty.
\end{equation}

\end{widetext} %%%%%%%%%%%%%%%%%%%%%%%%%%%%%%%%%%%%%%%%%%%%%%%%%%%

Employing asymptotic matching between Eqs.~(\ref{out}) and (\ref{in}), recalling that 
$\vec{\rho}=(\vec{r}-\vec{r}_1(t))/{\varepsilon}$, yields:
\begin{equation}
\notag
\frac{-2\vec{K}{+}\dot{\vec{r}}_1}{\ln \varepsilon-\ln(\psi_{\rm TF}(\vec{r}_1))}-\frac{2\nabla^\bot \psi_{\rm TF}(\vec{r}_1) }{\psi_{\rm TF}(\vec{r}_1)}=0.
\end{equation}
Thus, to leading order of $\nu\equiv {-1}/{\ln \varepsilon}$, we obtain
\begin{equation}\label{dynamics}
\dot{\vec{r}}_1 = {-}\frac{2}{\nu} \frac{\nabla^\bot \psi_{\rm TF}(\vec{r}_1) }{\psi_{\rm TF}(\vec{r}_1)} {+}2\vec{K}.
\end{equation}
Now, depending on the range of $|\vec{r}_1|$, we have two different leading order dynamics:
\begin{itemize}
\item 
If 
$|\nabla^\bot \psi_{\rm TF}(\vec{r}_1)| \ll \nu$: the dominant term in (\ref{dynamics}) is $\vec{K}$
and thus
\begin{equation}\label{dynamics1}
\dot{\vec{r}}_1= {2\vec{K}}
\end{equation}
\item 
If 
$ |\nabla^\bot \psi_{\rm TF}(\vec{r}_1)| \gg \nu$: the dominant term in (\ref{dynamics}) 
is the first term and thus
\begin{equation}
%\notag
\dot{\vec{r}}_1 ={-}\frac{2}{\nu} \frac{\nabla^\bot \psi_{\rm TF}(\vec{r}_1) }{\psi_{\rm TF}(\vec{r}_1)}
\end{equation}
\end{itemize}
While the comparison of the relevant cases in terms of
the dominant mathematical contribution is straightforward,
assigning an intuitive explanation to these different scenarios
is an open topic worthwhile of further consideration in future studies.
By returning to the original (unscaled) variable $t$,
Eq.~(\ref{dynamics1}) finally yields the following
expression for the rate of change
of the vortex position vector $\vec{r}_1$:
\begin{equation}
\notag
\dot{\vec{r}}_1 = \frac{\vec{K}}{R_{\rm TF}^2},
\end{equation}
which, by noting that $\vec{K}$ is an azimuthal vector, 
is in agreement with the precession frequency that was obtained using the
asymptotic expansion for the energy in Sec.~\ref{sec:sub:gral}:
\begin{equation}
\label{freq_append}
\omega 
%= \dot\phi_1  
\sim \frac{1}{R_{\rm TF}^2} .
\end{equation}
%

%%%%%%%%%%%%%%%%%%%%%%%%%%%%%%%%%%%%%%%%%%%%%%%%%%%%%%%%%%%%%%%%%%%%%%%%%%%%%%

\end{document}